\documentclass[12pt]{article}

\usepackage[totalheight = 23cm, totalwidth = 17cm]{geometry}
\usepackage{amssymb,amsmath,amsfonts,amsbsy}

\def\mathbi#1{\textbf{\em #1}}

\newcommand{\mpl}{m_{\rm Pl}}
\newcommand{\fnl}{f_{\rm NL}}
\newcommand{\calO}{{\cal O}}
\newcommand{\calP}{{\cal P}}
\newcommand{\calR}{{\cal R}}

\begin{document}

\begin{titlepage}

\rightline{\footnotesize{APCTP-Pre2014-005}}

\begin{center}

\vskip 1.0cm

\textbf{\Huge Blue running of the \\ primordial tensor spectrum}

\vskip 1.0cm

\large{
Jinn-Ouk Gong$^{a,b}$
}

\vskip 0.5cm

\small{\it 
$^{a}$Asia Pacific Center for Theoretical Physics, Pohang 790-784, Korea 
\\
$^{b}$Department of Physics, Postech, Pohang 790-784, Korea
}

\vskip 1.2cm

\end{center}

\begin{abstract}

We examine the possibility of positive spectral index of the power spectrum of the primordial tensor perturbation produced during inflation in the light of the detection of the $B$-mode polarization by the BICEP2 collaboration. We find a blue tilt is in general possible when the slow-roll parameter decays rapidly. We present two known examples in which a positive spectral index for the tensor power spectrum can be obtained. We also briefly discuss other consistency tests for further studies on inflationary dynamics.

\end{abstract}

\end{titlepage}

\setcounter{page}{0}
\newpage
\setcounter{page}{1}

\section{Introduction}

Cosmic inflation is considered as the leading candidate to resolve otherwise finely tuned initial conditions of the hot big bang cosmology such as the horizon problem~\cite{inflation}, and at the same time to provide the origin of large scale inhomogeneities~\cite{Mukhanov:1981xt}. Many properties of the seed perturbation, the curvature perturbation $\calR$ produced during inflation, are dramatically constrained by the observations on the temperature fluctuations of the cosmic microwave background (CMB). From the most recent Planck data, we have very good constraints on the amplitude of the power spectrum $\calP_\calR$, its spectral index $n_\calR$, and so on~\cite{Ade:2013zuv}. The predictions of the simplest chaotic inflation model~\cite{Linde:1983gd} are mostly consistent with the observations, but the expected tensor-to-scalar ratio $r \sim 0.15$ was in tension with the Planck constraint $r < 0.11$~\cite{Ade:2013uln}. Instead, $R^2$ inflation~\cite{Starobinsky:1980te} and related theories such as the standard model Higgs inflation~\cite{alstar_ext} have attracted a lot of attention, as $r$ is very small in these scenarios.

Recently, however, the BICEP2 experiment announced the detection of the $B$-mode polarization of the CMB in the range $30 < \ell < 150$. At 1$\sigma$ confidence level, the corresponding value of $r$ reads~\cite{Ade:2014xna}
\begin{equation}
r = 0.20^{+0.07}_{-0.05} \, ,
\end{equation}
with $r = 0$ disfavoured at 7$\sigma$. Subtracting foreground dust modifies the allowed range slightly, $r = 0.16^{+0.06}_{-0.05}$ but still $r = 0$ is strongly disfavoured at $5.9\sigma$. This is in good agreement with the prediction of simple chaotic inflation, indicating that the energy scale during inflation is as high as $10^{16}$ GeV.

Thus our next task should be, with more data available soon, to examine more closely the predictions of the models that survive the current constraints. In this regard, a tantalizing feature is that the BICEP2 $BB$ auto spectrum $C_\ell^{BB}$ exhibits a bump. This may well be due to lack of more data or systematics, and may disappear after the Keck Array data analysis is completed. This feature, however, could be real as well: in that case, the power spectrum of the primordial tensor perturbation has a {\em blue} tilt: $n_T \sim 0.50 \pm 0.25$. The spectrum of tensor perturbation may have a small blue tilt in some string theory motivated models~\cite{stringgas}, but in the standard inflationary scenario the tilt is given by $n_T = -2\epsilon$, with $\epsilon \equiv -\dot{H}/H^2 > 0$ being the slow-roll parameter, and thus appears to be always red (see however~\cite{Contaldi:2014zua}). In this article we examine the running of the power spectrum of tensor perturbation and explore the possibility of a positive $n_T$.

\section{Blue spectral index}

Our starting point is to recall a more accurate formula for the spectral index $n_T$ of the tensor power spectrum $\calP_T$. In the general slow-roll formalism~\cite{gsr}, with ``general'' in the sense that we generalize the standard assumption for the slow-roll approximation in such a way that the time derivatives of $\epsilon$ need not be smaller than $\epsilon$ itself, $n_T$ can be readily calculated by taking a logarithmic derivative of $k$ of $\calP_T$~\cite{Gong:2004kd}. Evaluated at the moment of horizon crossing, up to second order corrections it is given by
\begin{equation}\label{nT-gsr}
n_T = 2\frac{p'}{p} + 2\alpha \left( \frac{p'}{p} \right)' + 2 (4-\pi) \frac{p'p''}{p^2} \, ,
\end{equation}
where $p = p(\log{x}) \equiv 2\pi xa/k$ with $x \equiv -k\eta$, a prime denotes a derivative with respect to $\log{x}$, and $\alpha \equiv 2-\log2-\gamma \approx 0.729637$, with $\gamma \approx 0.577216$ being the Euler-Mascheroni constant. Requiring (\ref{nT-gsr}) be positive leads to the condition on the fundamental general slow-roll function $p$,
\begin{equation}\label{p-condition}
\frac{p'}{p} \gtrsim 1 \, .
\end{equation}
Once this condition is satisfied we can have a blue tensor spectral index.

It is illustrative to make use of the slow-roll parameters to understand what (\ref{p-condition}) means. Writing $p'/p$ in terms of the slow-roll parameters~\cite{Gong:2004kd}, we can easily find
\begin{equation}\label{condition2}
\epsilon \sim a^{-1} \, ,
\end{equation}
that is, $\epsilon$ is rapidly decaying proportional to $1/a$ or even faster. This means another widely used slow-roll parameter,
\begin{equation}\label{def:eta}
\eta \equiv \frac{\dot\epsilon}{H\epsilon} \, ,
\end{equation}
is negatively large. If this condition is satisfied, during that period $\eta$ contribution can be larger than the leading term $-2\epsilon$ and make $n_T > 0$. Of course this period cannot last too long time otherwise inflation does not terminate. We need to provide a mechanism which leads to a graceful exit, as is obvious from the $BB$ spectrum from BICEP2 where the bump seems to disappear on smaller scales.

Before closing this section, we note that in~\cite{Gong:2004kd} $n_T$ is presented in a form which can be evaluated at any convenient point of evaluation, denoted by $\star$, in such a way that
\begin{equation}\label{def:alphastar}
\alpha \to \alpha_\star \equiv \alpha - \log \left( -k\eta_\star \right) \, ,
\end{equation}
with the slow-roll parameters being evaluated at the same moment. One may doubt that the value of $n_T$ is different depending on when it is evaluated. But it is not: as can be read from the integral formula of $\calP_T$ in~\cite{Gong:2004kd} computed in the context of the general slow-roll formalism~\cite{gsr}, the power spectrum is independent of the evaluation point $\star$, so is its spectral index $n_T$. As we can see from (\ref{def:alphastar}), when we evaluate at the moment of horizon crossing, or more precisely $-k\eta_\star=1$ which is different from horizon crossing by $\calO(\epsilon)$, the logarithmic factor disappears and the calculation becomes most transparent.

\section{Working examples}

Now we consider two simple examples in which the condition for a blue spectral index $n_T$ can be accomplished. We however do not attempt extensive BICEP2 data analysis or realistic model building here, which are both beyond the scope of this short article. From (\ref{condition2}), we can see that typical examples include the models with very flat inflaton potential $V(\phi)$. Let us consider a canonical inflaton field evolving on a constant potential $V(\phi) = V_0$~\cite{usr}. We can solve trivially the equation of motion for the background inflaton field and find $\dot\phi \propto a^{-3}$, and thus the slow-roll parameters are given by
\begin{equation}
\begin{split}
\epsilon & \propto a^{-6} \, ,
\\
\eta & = -6 \, .
\end{split}
\end{equation}
Thus (\ref{condition2}) is satisfied, and the tensor power spectrum during this stage exhibits a blue tilt. This picture, however, is plagued by the graceful exit problem and we should provide a mechanism that leads to the end of inflation~\cite{usr,Namjoo:2012aa}.

Note that in this model, there are other interesting phenomenology. For example, the local non-linear parameter is given by~\cite{Namjoo:2012aa}
\begin{equation}
\fnl = \frac{5}{2} \, ,
\end{equation}
which interestingly violates the consistency relation for the bispectrum of the curvature perturbation in the squeezed configuration~\cite{ngconsistency}. There are other similar consistency relations which include tensor perturbation: for example, the three-point correlation function of two tensor perturbations $h_s(\mathbi{k})$ with polarization $s$ and one curvature perturbation in the squeezed limit reads~\cite{ngconsistency}
\begin{equation}
\left\langle h_s(\mathbi{k}_1)h_{s'}(\mathbi{k}_2)\calR(\mathbi{k}_3) \right\rangle \underset{k_3\to0}{\longrightarrow} (2\pi)^3 \delta^{(3)}(\mathbi{k}_1+\mathbi{k}_2+\mathbi{k}_3) P_T(k_1) P_\calR(k_3) n_T \delta_{ss'} \, ,
\end{equation}
where $P_i(k) \equiv 2\pi^2\calP_i(k)/k^3$ with $i=\calR$ and $T$.

In the other example the inflaton potential is given by~\cite{Jain:2008dw}
\begin{equation}
V(\phi) = \frac{1}{2}m^2\phi^2 - \frac{\sqrt{2\lambda(n-1)}}{n} m\mpl^3 \left( \frac{\phi}{\mpl} \right)^n + \frac{1}{4}\lambda\mpl^4 \left( \frac{\phi}{\mpl} \right)^{2(n-1)} \, ,
\end{equation}
with $n>2$ being an integer, which can arise in certain minimal supersymmetric extensions of the standard model~\cite{Allahverdi:2006iq}. This potential allows, depending on the combination of the three model parameters $m$, $\lambda$ and $n$, a brief period of departure from inflation sandwiched between two stages of slow-roll inflation. (\ref{condition2}) is satisfied during the intermediate stage and the tensor spectrum shows a blue tilt. However, as can be seen from the explicit illustrations of Figure~2 in~\cite{Jain:2009pm}, much more prominent is the enhancement in $r$ due to the suppression of $\calP_\calR$, which can be as large as $r\gg1$. For similar studies due to fast-roll phase, see e.g.~\cite{fastroll}.

\section{Discussions and conclusions}

In this article we have examined whether the power spectrum of the primordial tensor perturbation $\calP_T$ may have a non-trivial running $n_T$ in such a way that it is blue in some range as suggested by the observed BICEP2 $BB$ spectrum $C_\ell^{BB}$. This is possible when the slow-roll parameter $\eta$ is negatively large. Practically, however, we can only constrain $n_T$ as a whole, not the individual contribution. This is especially the case when we consider models, for example, with feature because the CMB power spectrum has strong degeneracies~\cite{Gong:2005jr}. An interesting but simple way to disentangle different contributions, in particular to eliminate $\epsilon$, is the running of $r$~\cite{Gong:2007ha}. For a few interesting cases, we find
\begin{equation}
\frac{d\log{r}}{d\log{k}} = 1-n_\calR+n_T = \left\{
\begin{array}{ll}
\eta & \text{ for canonical single field,}
\\
\eta+s & \text{ for general single field,}
\\
\eta_\text{multi} + 2 \dfrac{N_{,i}N_{,j}}{G^{kl}N_{,k}N_{,l}} \dfrac{R^i{}_{ab}{}^j}{3\mpl^2} \dfrac{\dot\phi^a\dot\phi^b}{H^2} & \text{ for multi-field,}
\end{array}
\right.
\end{equation}
where $c_s$ is the speed of sound which may be non-trivial~\cite{heavy}, $s \equiv \dot{c}_s/(Hc_s)$, $N_{,i}$ is the derivative of the $e$-fold $N$ with respect to $\phi^i$, $R^i{}_{jkl}$ is the Riemann curvature tensor constructed from the field space metric $G_{ij}$, and\footnote{Here we define $\eta_\text{multi}$ different from~\cite{Gong:2007ha}.}~\cite{deltaN}
\begin{equation}
\eta_\text{multi} = \frac{r}{4} - 2\mpl^2 \frac{N_{,i}N_{,j}}{G^{kl}N_{,k}N_{,l}} \frac{V^{;ij}}{V} \, ,
\end{equation}
with a semicolon denoting a covariant derivative in the field space. Combined with the consistency relation for the local $\fnl$, we can build a web of consistency checks and should be able to correlate one signal to another explicitly~\cite{corr-corr}.

To summarize, we have considered the general possibility $n_T > 0$ in the context of standard inflationary models. This can be accomplished if the slow-roll parameter $\epsilon$ is exponentially decaying during inflation. We have presented two simple examples where this condition is satisfied. Combined with other consistency tests including the running of $r$, we can further study various aspects of inflationary dynamics.

\subsection*{Acknowledgements}

I acknowledge the Max-Planck-Gesellschaft, the Korea Ministry of Education, Science and Technology, Gyeongsangbuk-Do and Pohang City for the support of the Independent Junior Research Group at the Asia Pacific Center for Theoretical Physics. I am also supported by a Starting Grant through the Basic Science Research Program of the National Research Foundation of Korea (2013R1A1A1006701).

\end{document}